\documentclass{article}

\usepackage[utf8]{inputenc}
\usepackage[small]{dgruyter}
\usepackage{microtype}

\usepackage{bm}
\usepackage{amsmath}
\usepackage{indentfirst}
\usepackage{graphicx}
\usepackage{subfigure}
\usepackage{authblk}
\usepackage{threeparttable}
\usepackage{booktabs}
\usepackage{hyperref}

\usepackage{multirow, color}
\usepackage{mathrsfs}
\usepackage{setspace}
\usepackage{bbm}
\usepackage[figuresright]{rotating}
\usepackage{ulem}
 \usepackage{mathrsfs}
 \usepackage{float}
\usepackage[T1]{fontenc}
\usepackage[utf8]{inputenc}
\usepackage{authblk}
\usepackage{breqn}
\usepackage{url}
\usepackage{changepage}
\usepackage{adjustbox}
\usepackage{verbatim}

\newtheorem{result}{Result}

\begin{document}

  \articletype{}
  \title{ Estimating the natural indirect effect and the mediation proportion via the product method}
  
  \author[1,2,*]{Chao Cheng}
  \author[1,2]{Donna Spiegelman}
  \author[1,2]{Fan Li} 
  \runningauthor{Cheng, Spiegelman and Li}
  \affil[1]{Department of Biostatistics, Yale School of Public Health, New Haven, CT}
  \affil[2]{Center for Methods in Implementation and Prevention Science, Yale School of Public Health, New Haven, CT}

  \runningtitle{Mediation analysis with the product method}
  \subtitle{}
\abstract{The natural indirect effect (NIE) and mediation proportion (MP) are two measures of primary interest in mediation analysis. The standard approach for estimating NIE and MP is through the product method, which involves an model for the outcome conditional on the mediator and exposure and another model describing the exposure--mediator relationship. The purpose of this article is to comprehensively develop and investigate the finite-sample performance of NIE and MP estimators via the product method. 
With four common data types, we propose closed-form interval estimators via the theory of estimating equations and multivariate delta method, and evaluate its empirical performance relative to the bootstrap approach. In addition, we have observed that the rare outcome assumption is frequently invoked to approximate the NIE and MP with a binary outcome, although this approximation may lead to non-negligible bias when the outcome is common. We therefore introduce the exact expressions for NIE and MP with a binary outcome without the rare outcome assumption and compare its performance with the approximate estimators. Based upon these theoretical developments and empirical studies, we offer several practical recommendations to inform practice. An R package \texttt{mediateP} is developed to implement the methods for point and variance estimation discussed in this paper.}

\maketitle

\noindent {\small *Correspondence to Chao Cheng, MS, Department of Biostatistics, Yale School of Public Health, 135 College Street, Ste 200, New Haven, CT 06510 (e-mail: c.cheng@yale.edu)}

\clearpage

\section{Introduction}

Biomedical and epidemiological studies evaluating the impact of exposure on health outcomes have received considerable attention in recent years. In addition to estimating the total effect of the exposure on the disease outcome, it is also of interest to explore potential pathways and mechanisms underlying these exposure-disease relationships. An important tool for exploring these pathways is mediation analysis \cite{baron1986moderator,vanderweele2015explanation}, which decomposes the \textit{total effect} (TE) of the exposure into two components, a \textit{natural indirect effect} (NIE) through a pre-specified intermediate variable (i.e., the mediator) and a \textit{natural direct effect} (NDE) whose impact derives solely from the exposure. Formal definitions of NIE and NDE were first given in \cite{robins1992identifiability,10.5555/2074022.2074073} under a causal framework. The NIE suggests how much we could exploit the exposure-disease mechanism through potential interventions targeting the mediator, especially in cases when it is difficult to manipulate the exposure \cite{barfield2017testing}.

In mediation analysis, researchers sometimes calculate the ratio of the NIE and the TE to capture the relative importance of the mediator in explaining the pathway through which the exposure has an effect on the outcome \cite{vanderweele2013policy}. This ratio is called \textit{mediation proportion} (MP) or equivalently, the \textit{proportional mediated}. There has been an increasing number of epidemiological studies that report the MP to explain the exposure-disease mechanism when conducting mediation analysis \cite{bebu2017relationship,bowe2020diabetes,chang2019associations,huang2017therapeutic,inzaule2018previous,parker2016prenatal}, and some of these studies include only a small to moderate sample size not exceeding 1000. Given the increasing use of MP, a comprehensive evaluation of the finite-sample operating characteristics of the point and interval estimators for MP is needed to better inform practice.

Many statistical methods have been developed to identify and estimate the natural indirect effect from observational data, including but not limited to the nonparametric approach \cite{imai2010general,imai2010identification}, the weighting-based semiparametric approach \cite{tchetgen2012semiparametric,tchetgen2014estimation,huber2015direct}, as well as the parametric outcome regression approach \cite{mackinnon1995simulation,vanderweele2015explanation,vanderweele2016mediation,nevo2017estimation}. This paper will focus on the parametric outcome regression approach. 
The parametric outcome regression approach includes two major variants, the \textit{difference method} \cite{nevo2017estimation,jiang2015difference} and the \textit{product method} \cite{vanderweele2015explanation}. The difference method evaluates two regression models for the outcome with and without conditioning on the mediator, whereas the product method evaluates a regression model for the outcome conditional on the exposure and mediator and another regression model for the mediator conditional on the exposure.
When both the outcome and mediator variables are continuous and the ordinary least-squares is used to estimate regression model parameters, the difference method and product method are algebraically equivalent \cite{mackinnon1995simulation} for estimating the mediation effects. With a binary outcome or a binary mediator, however, the product method generally does not coincide with the difference method, and both the product and difference methods could produce unbiased estimates of NIE and MP under their respective model assumptions. 

While the estimation and inference for the mediation measures via the difference method have been extensively studied \cite{baron1986moderator,mackinnon1995simulation,jiang2015difference,nevo2017estimation}, several important issues remain less clear when using the product method for mediation analysis. First, the bootstrap approach has been suggested by a number of authors to construct valid confidence intervals of the mediation measures \cite{bollen1990direct,mackinnon2007mediation,mackinnon2004confidence,shrout2002mediation}. The explicit expressions of the closed-form asymptotic variance and interval estimators and their empirical performance have not previously been sufficiently detailed. 
 Second, when the outcome is binary, traditional mediation analyses have make the rare disease or outcome assumption and estimate the approximate NIE and MP under that assumption \cite{vanderweele2015explanation,valeri2013mediation}. Several recent publications have proposed exact mediation estimators for a common binary outcome \cite{gaynor2019mediation,samoilenko2018comparing,doretti2021exact}, but the empirical performance of these exact estimators has not been extensively evaluated. Specifically, the relative performance of these exact estimators compared to the approximate estimators under the rare and common outcome scenarios has not been studied in detail, and not at all with a binary outcome and a continuous mediator.
Third, while the empirical performance of the NIE estimator has been investigated in prior studies \cite{mackinnon1995simulation,barfield2017testing,mackinnon2004confidence}, there has been relatively little empirical evaluations for the MP estimates via the product method. Given that the MP is of primary interest in epidemiology and medicine, a comprehensive empirical evaluation of the MP estimator based on the product method is valuable to inform practice.

 To address the aforementioned issues, we conducted an extensive Monte Carlo study to evaluate the point and interval estimators for NIE and MP under the four common data types: \textbf{Case \#1}, a continuous outcome and a continuous mediator; \textbf{Case \#2}, a continuous outcome and a binary mediator; \textbf{Case \#3}, a binary outcome and a continuous mediator; as well as \textbf{Case \#4}, a binary outcome and a binary mediator. We reviewed the counterfactual outcome framework for estimating the NIE and MP via the product method, and importantly,
derived the closed-form variance and interval estimators through the method of generalized estimating equations and the multivariate delta method. We provided a comparison between the closed-form interval estimator and the bootstrap interval estimators via simulations with different sample sizes. In the scenario with a binary outcome and a continuous mediator, we also developed the exact NIE and MP expressions, obviating the rare outcome assumption, and evaluate the performance of the exact versus approximate expressions under rare and common outcome assumptions. Thus, an important goal of our study is to unify and supplement existing evidence on the empirical performance of the product method, especially when closed-form variance is considered and when the rare outcome assumption fails to hold. To better elucidate our contribution to the literature, Table \ref{tab:summary} summarizes the scenarios considered in our study as compared to several prior studies \cite{barfield2017testing,biesanz2010assessing,fritz2007required,gaynor2019mediation,mackinnon1995simulation,mackinnon2002comparison,mackinnon2004confidence,rijnhart2019comparison,samoilenko2018comparing}, across the four common data types.


The remainder of this paper is organized as follows. In Section \ref{sec:methods}, we describe the product method to obtain the point and interval estimators of NIE and MP in the aforementioned four cases with different data types. In Section \ref{sec:sim}, we describe our Monte Carlo simulation study that investigates the performance of the point and interval estimators for NIE and MP, and report our findings. To illustrate the product method, we study in Section \ref{sec:app} how much the effect of an anti-retroviral delivery intervention on 12-month retention is mediated by the 6-month visit adherence in the MaxART study \cite{khan2020early}. Section \ref{sec:discussion} concludes with a brief discussion. 

\section{Mediation analysis via the product method}\label{sec:methods}

\subsection{Mediation measures and the product method}

\begin{figure}[htp]
\centering 
\includegraphics[width = 8 cm]{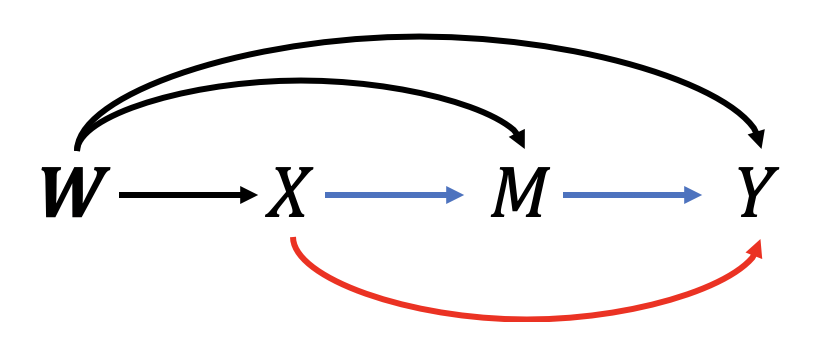}
\caption{Mediation directed acyclic graph, where $Y$, $X$, $M$ and $\bm W$ denote the outcome, exposure, mediator, and confounders of the exposure-outcome and exposure-mediator relationships. The NIE of exposure $X$ on outcome $Y$ through mediator $M$ is highlighted in blue and the NDE of exposure $X$ on outcome $Y$ is highlighted in red.  }
\label{fig:causalgraph}
\end{figure}

Assume that we have an outcome of interest, $Y$, an exposure, $X$, and a mediator, $M$, where each variable can be continuous or binary. We also observe $\bm W$, a vector of covariates, associated with outcome and measured before the exposure, some of which may be confounders of the estimated exposure-outcome association and/or the mediator-outcome association.  A directed acyclic graph illustrating the causal relationship between those variables is shown in Figure \ref{fig:causalgraph}. We are interested in identifying two causal effects, the NIE ad NDE. To identify these causal effects, we will follow the counterfactual framework used in the classic causal inference literature \cite{robins1992identifiability,10.5555/2074022.2074073}. Specifically, we will follow the notation in \cite{nevo2017estimation} and define $Y_x$ and $M_x$ as the outcome $Y$ and mediator $M$, respectively, that would have been observed when setting $X=x$. Similarly, let $Y_{x,m}$ be the  value of outcome $Y$ that would have been observed when setting $X=x$ and $M=m$.

\newcommand\independent{\protect\mathpalette{\protect\independenT}{\perp}}
\def\independenT#1#2{\mathrel{\rlap{$#1#2$}\mkern2mu{#1#2}}}

Based on classical counterfactual frameworks, we make the following assumptions. First is the consistency assumption, which assumes that the potential outcome $Y_{x}$ and potential mediator $M_x$ equal their respective observed variables $Y$ and $M$ if we set $X=x$ \cite{vanderweele2009conceptual}. Similarly, we also require that $Y_{x,m}$ equals $Y$ if $X=x$ and $M=m$ are observed \cite{vanderweele2010odds}. The second is the composition assumption, which requires that $Y_{x}=Y_{x,M_x}$, i.e., the potential outcome $Y$ where $X=x$ is equal to the potential outcome $Y$ when $X=x$ and $M$ is set to its value corresponding to when $X=x$ \cite{vanderweele2009conceptual}. In order to identify NIE and NDE, we also require several identification assumptions in relation to confounding, including (A.1) $Y_{x,m} \independent X | \bm W$, (A.2) $Y_{x,m} \independent M | \bm W, X$, (A.3) $M_x \independent X | \bm W$, and (A.4) $Y_{x,m} \independent M_{x^*} | \bm W$ for all $x$, $x^*$, and $m$. That is, (A.1)--(A.3) assume that the exposure-outcome, mediator-outcome, and exposure-mediator relationships are not confounded conditional on covariates $\bm W$. (A.4) is sometimes termed cross-world independence, which stresses that none of the confounders in the mediator-outcome relationship can be affected by exposure $X$ \cite{doretti2021exact}.
 
Under this framework, when changing the exposure levels from $x^*$ to $x$ conditional on $\bm W=\bm w$, Nevo et al. \cite{nevo2017estimation} defined the NIE and NDE on a $g$-function scale:
\begin{equation}\label{causalframe}
\begin{aligned}
& \text{NIE}(x^*,x|\bm w) = g\left(E[Y_{x,M_x}|\bm W=\bm w]\right) - g\left(E[Y_{x,M_{x^*}}|\bm W=\bm w]\right), \\
& \text{NDE}(x^*,x|\bm w) = g\left(E[Y_{x,M_{x^*}}|\bm W=\bm w]\right) - g\left(E[Y_{x^*,M_{x^*}}|\bm W=\bm w]\right),
\end{aligned}
\end{equation}
where $g(.)$ is a pre-specified monotone function. In this article, $g(.)$ is set to the link function of the model for conditional mean $Y$, which will be discussed later. Given \eqref{causalframe}, the TE is defined as summation of NIE and NDE, that is,
$$
\text{TE}(x^*,x|\bm w) = g\left(E[Y_{x,M_{x}}|\bm W=\bm w]\right) - g\left(E[Y_{x^*,M_{x^*}}|\bm W=\bm w]\right). \\
$$
Finally, the MP is given by the ratio of NIE and TE.

The above mediation measures are derived while fixing $\bm W$ to $\bm w$ while changing the exposure level from $x^*$ to $x$, and therefore are conditional causal parameters. Alternatively, definitions of mediation measures have been proposed by averaging over the distribution of $\bm W$. As discussed in \cite{imai2010general,imai2010identification}, under assumptions (A.1)--(A.4), the NIE, NDE, and TE are given by $\text{NIE}(x^*,x) = E[Y_{x,M_x}] - E[Y_{x,M_{x^*}}]$, $\text{NDE}(x^*,x) = E[Y_{x,M_{x^*}}] - E[Y_{x^*,M_{x^*}}]$, and $\text{TE}(x^*,x) = E[Y_{x,M_{x}}] - E[Y_{x^*,M_{x^*}}]$, respectively. Estimation and inference for the marginal mediation measures has been discussed in \cite{imai2010identification,pearl2012causal,tchetgen2012semiparametric}, based on nonparametric or semiparametric models for the outcome and mediator, as commonly appear in practice, and will not be pursued here. The focuse of this paper is estimation and inference mediation measures conditional on $\bm W = \bm w$, as shown in \eqref{causalframe}. 

We now introduce the product method for estimating those mediation effects. Specifically, we assume the following conditional mean model for the  outcome ($Y$),
\begin{equation}\label{outcome1}
g(E(Y|X,M,\bm W)) = \beta_0 + \beta_1 X + \beta_2 M + \bm \beta_3^T \bm W,
\end{equation}
where $g(.)$ is a link function, $\beta_1$ is the exposure effect on the outcome conditional on the effects of the mediator and confounders, $\beta_2$ represents the relationship between the mediator and outcome conditional on the effect of the exposure and confounders. Common link functions include the identity function when the outcome is continuous and logistic function when the outcome is binary.  Because previous empirical evidence suggests that there are few  interaction effects between an intervention/exposure and covariates 
 that replicated across studies in public health and epidemiology \cite{spiegelman2017evaluating}, we assume there are no mediator--exposure interactions in \eqref{outcome1}. Mediation analyses in the presence of mediator--exposure interaction effects are studied elsewhere, for example, \cite{valeri2013mediation,vanderweele2010odds,gaynor2019mediation}. Researchers can empirically verify this assumption in their data before applying these methods.

In addition to the outcome model (\ref{outcome1}), the product method additionally requires fitting the following model for the mediator:
\begin{equation}\label{mediator1}
h\left(E(M|X,\bm W)\right) = \gamma_0 + \gamma_1 X + \bm \gamma_2^T \bm W,
\end{equation}
where $\gamma_1$ represents the association between the exposure to the mediator conditional on the effects of the covariates, $h$ is a link function, which can be a identity function and a logistic function when the mediator is continuous and binary, respectively.  For simplicity of notation but without loss of generality, we assume that the mediator model and the outcome model share the same set of covariates. We can set some elements in $\bm \beta_3$ or $\bm \gamma_2$ to zero when the covariate sets in the outcome and mediator models are not exactly the same.

Next, we provide expressions of the mediation measures under the scenarios of continuous and binary mediator, separately.

\subsection{Continuous mediator}\label{sec:cont}

 
When the mediator is continuous and $h$ is an identity link function, model \eqref{mediator1} becomes
\begin{equation}\label{contmediator}
 E(M|X,\bm W) = \gamma_0 + \gamma_1 X + \bm \gamma_2^T \bm W.
\end{equation}
First consider Case \#1, where the outcome is also continuous and $g$ is an identity link function. As indicated by \cite{valeri2013mediation}, if the identification assumptions hold and the outcome as well as the mediator models are correctly specified, the NIE, NDE and TE can be expressed as $\beta_2\gamma_1(x-x^*)$, $\beta_1(x-x^*)$ and $(\beta_2\gamma_1+\beta_1)(x-x^*)$, respectively.  The mediation proportion is given as $\frac{\beta_2\gamma_1}{\beta_2\gamma_1+\beta_1}$. 

When the outcome is binary and $g(.)$ is the logistic link function (Case \#3), the NIE, NDE, and TE can be defined on the log odds ratio scale. In Web Appendix A, we derive exact expressions of mediation measures under the conditions that $M|X,\bm W$ in model \eqref{contmediator} follows a normal distribution with a constant variance $\sigma^2$. Specifically, define an integrand
$$\tau(x', x, m, \bm w)=\frac{\exp\{(\gamma_0+\gamma_1x'+\bm \gamma_2^T\bm w)m/\sigma^2-m^2/(2\sigma^2)\}}{1+\exp(\beta_0+\beta_1 x + \beta_2m + \bm \beta_3^T \bm w)} $$
we have that 
\begin{align}
\text{NIE}=&\log\left\{\frac{ \int_m  \exp(\beta_2 m)\tau(x,x,m,\bm w) \text{d}m  }{  \int_m  \tau(x,x,m,\bm w) \text{d}m  }\right\} - \log\left\{\frac{ \int_m  \exp(\beta_2 m)\tau(x^*,x,m,\bm w)\text{d}m  }{  \int_m \tau(x^*,x,m,\bm w) \text{d}m  }\right\},\label{eq:NIE3}\\
\text{NDE}=&\beta_1(x-x^*)+\nonumber\\
&\log\left\{\frac{ \int_m  \exp(\beta_2 m)\tau(x^*,x,m,\bm w)\text{d}m  }{  \int_m \tau(x^*,x,m,\bm w) \text{d}m  }\right\}-\log\left\{\frac{ \int_m  \exp(\beta_2 m)\tau(x^*,x^*,m,\bm w)\text{d}m  }{  \int_m \tau(x^*,x^*,m,\bm w) \text{d}m  }\right\}\label{eq:NDE3},
\end{align}
both of which involve one-dimensional logistic-normal integrals that do not have closed-form solutions. Gaynor et al. (2019) \cite{gaynor2019mediation} uses a probit function to approximate the logistic function in the integral and obtained closed-form expressions for the mediation measures (Table \ref{tab:expression}). However, the probit approximation tends to be inaccurate as the outcome prevalence  deviates from 50\%, as discussed in \cite{gaynor2019mediation} and Web Appendix A. Instead of using probit function to approximate the logistic-normal integrals, we consider in this paper the Gauss-Hermite Quadrature (GHQ) approach \cite{liu1994note} to numerically calculate the integrals. 


Provided the outcome is rare, we can approximate the NIE, NDE, and TE as $\beta_2\gamma_1(x-x^*)$, $\beta_1(x-x^*)$, and $(\beta_2\gamma_1+\beta_1)(x-x^*)$, respectively, as given by \cite{vanderweele2010odds,valeri2013mediation}. It follows that $\text{MP}\approx \frac{\beta_2\gamma_1}{\beta_2\gamma_1+\beta_1}$. Here we can obtain simple and closed-form expressions for the mediation measures because the logistic link function in model \eqref{outcome1} approximates a log link function, and the log binomial model is collapsible. Therefore, the logistic-normal integrals in the mediation expressions can be approximated by the log-normal integrals that have closed-form solutions \cite{vanderweele2010odds}. In Web Appendix A, we provide more details on the validity of the approximate expressions when the outcome is rare. In order to distinguish the approximate expressions under a rare outcome assumption from the exact expressions, we write those approximate mediation measures as $\text{NIE}^{(a)}$, $\text{NDE}^{(a)}$, $\text{TE}^{(a)}$, and $\text{MP}^{(a)}$, respectively. 

\subsection{Binary Mediator}\label{sec:bin}

In this section, we describe the product method estimators when $M$ is a binary variable, based on the following logistic regression model for $M$:
\begin{equation}\label{mediator2}
\text{logit}\Big(P(M=1|X,\bm W)\Big) = \gamma_0 + \gamma_1X+\bm \gamma_2^T \bm W.
\end{equation}
First, consider the case when outcome is continuous (Case \#2) and $g(.)$ is the identity link function. Throughout this Section, we define the exponentiated linear component $\kappa(x,\bm w)=\exp(\gamma_0+\gamma_1 x + \bm\gamma_2^T \bm w)$. If the identification assumptions (A.1)--(A.4) hold and the outcome as well as mediator models are correctly specified, we have
$$
\text{NIE} = \beta_2 \left\{ \frac{\kappa(x,\bm w)}{ 1+ \kappa(x,\bm w)} - \frac{\kappa(x^*,\bm w)}{ 1+ \kappa(x^*,\bm w)} \right\}=\beta_2\left\{\frac{\kappa(x,\bm w)-\kappa(x^*,\bm w)}{(1+ \kappa(x,\bm w))(1+ \kappa(x^*,\bm w))}\right\}$$
and $\text{NDE} = \beta_1 (x-x^*)$, as shown in \cite{barfield2017testing}.
As a result, the mediation proportion is given by $\text{MP} = \frac{\text{NIE}}{\text{NDE}+\text{NIE}} $.

Finally, consider a binary mediator and a binary outcome (Case \#4), where we fit the mediator model (\ref{mediator2}) and outcome model (\ref{outcome1}) with logistic link functions.  Then, as long as the identification assumptions (A.1)--(A.4) hold and (\ref{outcome1}) and (\ref{mediator2}) are correctly specified, the NIE and NDE on a log odds ratio scale is given by \cite{samoilenko2018comparing,doretti2021exact}:
\begin{align}
\text{NIE}=&\log\left\{\frac{1+e^{\beta_2}\eta(x,\bm w)+\kappa(x^*,\bm w)(1+\eta(x,\bm w))}
{1+e^{\beta_2}\eta(x,\bm w)+\kappa(x,\bm w)(1+\eta(x,\bm w))}\right\}
-\nonumber\\
&\log\left\{\frac{1+e^{\beta_2}\eta(x,\bm w)+e^{\beta_2}\kappa(x,\bm w)(1+\eta(x,\bm w))}
{1+e^{\beta_2}\eta(x,\bm w)+e^{\beta_2}\kappa(x^*,\bm w)(1+\eta(x,\bm w))}\right\},\label{eq:NIE4}\\
\text{NDE}=&\beta_1(x-x^*)+\log\left\{\frac{1+e^{\beta_2}\eta(x^*,\bm w)+\kappa(x^*,\bm w)(1+\eta(x^*,\bm w))}
{1+e^{\beta_2}\eta(x,\bm w)+\kappa(x^*,\bm w)(1+\eta(x,\bm w))}\right\}
+\nonumber\\
&\log\left\{\frac{1+e^{\beta_2}\eta(x,\bm w)+e^{\beta_2}\kappa(x^*,\bm w)(1+\eta(x,\bm w))}
{1+e^{\beta_2}\eta(x^*,\bm w)+e^{\beta_2}\kappa(x^*,\bm w)(1+\eta(x^*,\bm w))}\right\},\label{eq:NDE4}
\end{align}
where $\eta(x,\bm w)=\exp(\beta_0+\beta_1 x+\beta_3^T \bm w)$.
Given NIE and NDE, the MP is given by $\frac{\text{NIE}}{\text{NIE+NDE} }$.  If the outcome is rare, the following approximate NIE and NDE have been widely used \cite{vanderweele2015explanation}: 
$$\text{NIE}^{(a)} = \log\left\{  \frac{ (1+\kappa(x^*,\bm w))(1+e^{\beta_2}\kappa(x,\bm w))  }{ (1+\kappa(x,\bm w))(1+e^{\beta_2}\kappa(x^*,\bm w)) } \right\},$$ 
and $\text{NDE}^{(a)}=\beta_1 (x-x^*)$. As a result, $\text{MP}^{(a)}=\frac{\text{NIE}^{(a)}}{\text{NIE}^{(a)}+\text{NDE}^{(a)} }$.
When the outcome is rare, we provide an explanation that the approximate expressions are valid by exploiting the similarity between the logistic and log link functions (see Web Appendix B).


\subsection{Point and interval estimates for NIE and MP}\label{sec:ee}


In the previous two sections, we provided expressions for NIE and MP in Cases \#1--4, which are functions of the unknown parameters in the outcome model $\bm \beta = [\beta_0,\beta_1,\beta_2,\bm \beta_3^T]^T$, and unknown parameters in the mediator model, $\bm \gamma=[\gamma_0,\gamma_1,\bm \gamma_2^T]^T$. Exact mediation expressions for mediation measures in Case \#3 also involve the variance of the error term in the mediator model \eqref{mediator2}, $\sigma^2$. Let $\bm \theta$ denote all unknown parameters used in the expressions of NIE and MP, which is $[\bm \beta^T, \bm \gamma^T, \sigma^2]^T$ for the exact expressions in Case \#3 and $[\bm \beta^T, \bm \gamma^T]^T$ for other cases. Hereafter, we will rewrite the expressions of NIE and MP as $\mathscr{NIE}(\bm \theta)$ and $\mathscr{MP}(\bm \theta)$, respectively, to emphasize that those expressions are functions of $\bm \theta$.

In practice, the coefficients in the outcome model, $\hat{\bm \beta}$, can be obtained by solving the following generalized estimating equation (GEE) \cite{liang1986longitudinal}:
$$
U(\bm \beta) = \sum_{i=1}^n U_i(\bm \beta) = \sum_{i=1}^n \frac{\partial E(Y_i|X_i,M_i,\bm W_i)}{\partial \bm \beta} V_i^{-1} \Big(Y_i - E(Y_i|X_i,M_i,\bm W_i)\Big)=\bm 0,
$$
where $\{Y_i,X_i,M_i,\bm W_i\}_{i=1}^n$ are $n$ observations of $\{Y,X,M,\bm W\}$ and $V_i$ is a  working variance term for the outcome $Y_i$. The optimal asymptotic efficiency of $\hat{\bm \beta}$ will be obtained when $V_i = \text{Var}(Y_i)$. When $V_i$ is misspecified, the resultant $\hat{\bm \beta}$ is still consistent under mild regularity conditions but could be less efficient \cite{wang2003working}. Similarly, the coefficients in the mediator model, $\hat{\bm \gamma}$, can be obtained by solving the GEE below
$$
U(\bm \gamma) =\sum_{i=1}^n U_i(\bm \gamma) = \sum_{i=1}^n \frac{\partial E(M_i|X_i,\bm W_i)}{\partial \bm \gamma} V_i^{*-1} \Big(M_i - E(M_i|X_i,\bm W_i)\Big)=\bm 0,
$$
where $V_i^*$ is the working variance for $M_i$. Also, misspecification of $V_i^*$ only impacts the efficiency of $\hat{\bm \gamma}$. When evaluating the exact expressions in Case \#3, $\hat \sigma^2$ is needed and can be estimated by solving the GEE
$$
U(\sigma^2) =\sum_{i=1}^n U_i(\sigma^2) = \sum_{i=1}^n \left\{  \sigma^2 - \left(M_i - E(M_i|X_i,\bm W_i)\right)^2 \right\} = 0.
$$
After obtaining $\hat{\bm \theta}$ through the above estimating equations, we can calculate $\widehat{\text{NIE}}$ and $\widehat{\text{MP}}$ by plugging those parameter estimates in their expressions introduced in the previous sections; i.e., $\widehat{\text{NIE}}=\mathscr{NIE}(\hat{\bm \theta})$ and $\widehat{\text{MP}}=\mathscr{MP}(\hat{\bm \theta})$.  In what follows, we develop the closed-form asymptotic variance expressions of $\widehat{\text{NIE}}$ and $\widehat{\text{MP}}$ based on the multivariate delta method, and review the nonparametric bootstrap approach to obtain the confidence interval estimators. 

The multivariate delta method first estimates the variance-covariance matrix of $\hat{\bm \theta}$, abbreviated by $\hat{\bm \Sigma}_{\bm \theta}$, based on the theory of estimating equations \cite{liang1986longitudinal}. Specifically, $\text{Var}(\hat{\bm \beta})$, $\text{Var}(\hat{\bm \gamma})$, and $\text{Var}(\hat{\sigma}^2)$ can be estimated by the robust sandwich variance estimators \cite{liang1986longitudinal} based on their respective estimating equations. If the working variance terms, $V_i$ and $V_i^*$, are correctly specified when estimating $\hat{\bm \beta}$ and $\hat{\bm \gamma}$, one can also approximate $\text{Var}(\hat{\bm \beta})$ and $\text{Var}(\hat{\bm \gamma})$ by the negative inverse information matrix, $-\left\{\frac{\partial U(\bm \hat{\bm \beta})}{\partial \bm \beta^T}\right\}^{-1}$ and $-\left\{\frac{\partial U(\hat{\bm \gamma})}{\partial \bm \gamma^T}\right\}^{-1}$, respectively. In general, to obtain $\hat{\bm \Sigma}_{\bm \theta}$, we also need to estimate the covariances between $\hat{\bm \beta}$, $\hat{\bm \gamma}$, and $\hat \sigma^2$. However, as we state below, these asymptotic covariances are 0 since the estimating equations are asymptotically uncorrelated. We formalize this result below.
\begin{result}\label{res1}
Estimators $\hat{\bm \beta}$, $\hat{\bm \gamma}$, and $\hat \sigma^2$ obtained by solving $U(\bm \beta)=\bm 0$, $U(\bm \gamma)=\bm 0$, and $U(\sigma^2)=0$ are asymptotically uncorrelated and have zero asymptotic covariances.
\end{result}
Detailed proof of Result \ref{res1} is provided in Web Appendix C. Following this result, we have that $\hat{\bm \Sigma}_{\bm\theta} = \left[\begin{matrix} \widehat{\text{Var}}(\hat{\bm \beta}) & \bm 0 & \bm 0\\ 
\bm 0 & \widehat{\text{Var}}(\hat{\bm \gamma}) & \bm 0\\
\bm 0 & \bm 0 & \widehat{\text{Var}}(\hat{\sigma}^2) \\
\end{matrix}\right]$ for the exact expressions in Case \#3 and $\hat{\bm \Sigma}_{\bm\theta} = \left[\begin{matrix} \widehat{\text{Var}}(\hat{\bm \beta}) & \bm 0\\ 
\bm 0 & \widehat{\text{Var}}(\hat{\bm \gamma}) \\
\end{matrix}\right]$ for the remaining Cases. Finally, the variances of $\widehat{\text{NIE}}$ and $\widehat{\text{MP}}$ is obtained through the multivariate delta method \cite{oehlert1992note}:
\begin{equation*}
\begin{array}{c}
\widehat{\text{Var}}(\widehat{\text{NIE}}) =  \Big(\frac{\partial \mathscr{NIE}(\bm \theta) }{ \partial \bm \theta } \Big|_{\bm \theta=\hat{\bm \theta}}\Big)^T \widehat{\bm\Sigma}_{\bm \theta}  \frac{\partial \mathscr{NIE}(\bm \theta) }{ \partial \bm \theta } \Big|_{\bm \theta=\hat{\bm \theta}}\text{ and } \\
\widehat{\text{Var}}(\widehat{\text{MP}}) =  \Big(\frac{\partial \mathscr{MP}(\bm \theta) }{ \partial \bm \theta }\Big|_{\bm \theta=\hat{\bm \theta}} \Big)^T \widehat{\bm \Sigma}_{\bm \theta}  \frac{\partial \mathscr{MP}(\bm \theta) }{ \partial \bm \theta }\Big|_{\bm \theta=\hat{\bm \theta}}.
\end{array}
\end{equation*}
Given these variance estimators, the 95\% confidence intervals of NIE and MP can be computed by normal approximation as $\Big\{\widehat{\text{NIE}} - 1.96 \times \sqrt{\widehat{\text{Var}}(\widehat{\text{NIE}})}, \widehat{\text{NIE}} + 1.96 \times \sqrt{\widehat{\text{Var}}(\widehat{\text{NIE}})}\Big\}$ and $\Big\{\widehat{\text{MP}} - 1.96 \times \sqrt{\widehat{\text{Var}}(\widehat{\text{MP}})}, \widehat{\text{MP}} + 1.96 \times \sqrt{\widehat{\text{Var}}(\widehat{\text{MP}})}\Big\}$, respectively.

Alternatively, the nonparametric percentile bootstrap approach \cite{davison1997bootstrap,efron1994introduction} can be used to approximate the empirical distributions of $\widehat{\text{NIE}}$ by resampling the dataset with replacement and re-estimating all model parameters. The values for NIE are then calculated for each bootstrap dataset. This step of resampling and calculating NIE is repeated for a large of times (e.g., 1,000 times), and then the bootstrap distribution of NIE is obtained from the collection of estimates based on resampled datasets. Finally, the percentile bootstrap approach employs the 2.5\% and 97.5\% percentiles of the bootstrap sample distribution to obtain a 95\% confidence interval of NIE. Similarly, the percentile bootstrap approach can be used to obtain  a 95\% confidence interval of MP. When implementing the product method, several studies \cite{bollen1990direct,mackinnon2007mediation,mackinnon2004confidence,shrout2002mediation} suggested using the bootstrap approach to calculate confidence intervals for the mediation measures, because the finite-sample distribution of the mediation measure estimators, especially the MP estimator, can be skewed and may not be well approximated by the normal distribution. 
In what follows, we will compare the empirical performance between the closed-form sandwich confidence interval estimator and the bootstrap interval estimator, and assess when each approach provides accurate empirical coverage for all 4 Cases. 



\section{Simulation study}\label{sec:sim}

We conducted simulation studies under a range of scenarios likely to be encountered in practice to assess the performance of the point and interval estimators of NIE and MP. 
We considered the four data types introduced earlier. We considered 4 levels of sample sizes at 150, 500, 1,000, and 5,000 for the continuous outcome cases (Cases \#1 and \#2). With a baseline outcome prevalence of 3\%, to obtain  15, 30, 150, 600 expected disease cases, we considered sample sizes of 500, 1,000, 5,000, 20,000 for the binary outcome scenarios (Cases \#3 and \#4). For each data type and sample size, we considered the exposure ($X$) as a binary variable and for simplicity assumed that there were no confounders in both the mediator and outcome models. When the outcome was continuous, we set TE $\in (0.25, 0.5, 1)$, indicating small, medium and large total effects; otherwise, we set the TE to $(\log(1.2), \log(1.5), \log(2))$. For each value of TE, we considered MP $\in (0.05, 0.2, 0.5)$. All the mediation measures are defined for  a change in $X$ from 0 to 1. We conducted a factorial design of all combinations of TEs, MPs, and sample sizes. For each set of these factors, 5,000 replications were performed. The exact parameter constellations are summarized below. 

\textbf{Case \#1, continuous outcome and continuous mediator.} We first generated the exposure $X\sim \text{Bernoulli}(0.5)$. Then, we simulated the mediator $M\sim N(\gamma_0+\gamma_1 X, 1)$, where $\gamma_0=0$. $\gamma_1$ was chosen to 0.408, corresponding to the exposure-mediator correlation, $\text{Corr}(X,M)=0.2$. Finally, given $X$ and $M$, we simulated $Y\sim N(\beta_0 + \beta_1 X + \beta_2 M, 1)$, where $\beta_0$ was fixed as 0. We let $\beta_1=(1-\text{MP})\times \text{TE}$ and $\beta_2=\frac{\text{MP}\times\text{TE}}{\gamma_1}$ based on the relationships that  $\text{NDE}=(1-\text{MP})\times \text{TE}=\beta_1$ and $\text{NIE}=\text{MP}\times\text{TE}=\beta_2\gamma_1$.

\textbf{Case \#2, continuous outcome and binary mediator.} First, we simulated $X\sim \text{Bernoulli}(0.5)$. Then, we generated $M$ conditional on $X$ by the logistic regression $\text{logit}(P(M=1|X))=\gamma_0 + \gamma_1 X$. Noting that $\gamma_0=\log\left(\frac{P(M=1|X=0)}{1-P(M=1|X=0)}\right)$, we chose the values of $\gamma_0$ such that the baseline prevalence of the mediator $P(M=1|X=0)=0.2$. Similar to Case \#1, we chose $\gamma_1=0.903$ to give an exposure-mediator correlation $0.2$. Finally, we simulated $Y\sim N(\beta_0 + \beta_1 X + \beta_2 M, 1)$, where $\beta_0=0$, $\beta_1=(1-\text{MP})\times \text{TE}$ from the difinition $\text{NDE}=(1-\text{MP})\times \text{TE}=\beta_1$, and $\beta_2 = {\text{MP}\times \text{TE}}/\left\{\frac{e^{\gamma_0+\gamma_1  }}{ 1+ e^{\gamma_0+\gamma_1  }} - \frac{e^{\gamma_0 }}{ 1+ e^{\gamma_0}}\right\}$ by noticing $\text{MP}\times \text{TE} = \text{NIE} = \beta_2 \left\{\frac{e^{\gamma_0+\gamma_1  }}{ 1+ e^{\gamma_0+\gamma_1  }} - \frac{e^{\gamma_0 }}{ 1+ e^{\gamma_0}}\right\}$.

\textbf{Case \#3, binary outcome and continuous mediator.} We first generated $X$ and $M$ using the same procedure in Case \#1. Now, given $X$ and $M$, we used the following logistic regression model to simulate $Y$,
\begin{equation}\label{logistic_sim}
\text{logit}\Big(P(Y=1|X,M)\Big) = \beta_0 + \beta_1 X +\beta_2 M.
\end{equation}
Since $\beta_0=\log\left(\frac{P(Y=1|X=0,M=0)}{1-P(Y=1|X=0,M=0)}\right)$, we selected $\beta_0$ so the baseline outcome prevalence was 3\%. Then, we selected $\beta_1$ and $\beta_2$ by numerically solving the system of equations $\text{MP}\times \text{TE} = \mathscr{NIE}(\beta_0,\beta_1,\beta_2,\gamma_0,\gamma_1)$ and $(1-\text{MP})\times \text{TE}  = \mathscr{NDE}(\beta_0,\beta_1,\beta_2,\gamma_0,\gamma_1)$, where  $\mathscr{NIE}(\beta_0,\beta_1,\beta_2,\gamma_0,\gamma_1)$ and $\mathscr{NDE}(\beta_0,\beta_1,\beta_2,\gamma_0,\gamma_1)$ refer to the exact expressions of NIE and NDE, given in \eqref{eq:NIE3} and \eqref{eq:NDE3}.


\textbf{Case \#4, binary outcome and binary mediator.} We first generated $X$ and $M$ using the same procedure in Case \#2, then generated the outcome $Y$ using the logistic regression model (\ref{logistic_sim}). The values of $\beta_0$, $\beta_1$ and $\beta_2$ were obtained as follows.  We chose $\beta_0$ such that the baseline outcome prevalence is around 3\%. We then chose $\beta_1$ and $\beta_2$ by solving the system of equations $\text{MP}\times \text{TE} = \mathscr{NIE}(\beta_0,\beta_1,\beta_2,\gamma_0,\gamma_1)$ and $(1-\text{MP})\times \text{TE}  = \mathscr{NDE}(\beta_0,\beta_1,\beta_2,\gamma_0,\gamma_1)$, where  $\mathscr{NIE}(\beta_0,\beta_1,\beta_2,\gamma_0,\gamma_1)$ and $\mathscr{NDE}(\beta_0,\beta_1,\beta_2,\gamma_0,\gamma_1)$ refer to the exact expressions of NIE and NDE, given in \eqref{eq:NIE4} and \eqref{eq:NDE4}.

For each data type, we obtained $\widehat{\text{NIE}}$ and $\widehat{\text{MP}}$ by solving the estimating equations in Section \ref{sec:ee}, variance estimates of $\widehat{\text{NIE}}$ and $\widehat{\text{MP}}$ by the multivariate delta method, and 95\% confidence intervals of NIE and MP by the multivariate delta method and percentile bootstrap approach. 
We did not use the bootstrap approach for the sample size of 20,000 in Cases \#3 and \#4, due to the computational cost of performing the number of bootstrapping samples with a large dataset across 5,000 replications.  With a binary outcome, we first evaluated the performance of the estimators based on the approximate expressions (i.e., $\text{NIE}^{(a)}$ and $\text{MP}^{(a)}$), and compared the performance between the approximate and the exact estimators for the mediation measures by varying the baseline outcome prevalence from 1\% to 50\%, in Section \ref{sec:comp}.



The percent bias (Bias(\%)) was used to evaluate the accuracy of the point estimates of NIE and MP. It was calculated as the median bias relative to the true value over 5,000 replications.
We employed the ``median'' rather than the ``average'' value over all the replications to avoid the undue influence of outliers on the results when the sample sizes were not large. The variance ratio was defined as the the ratio between the median of the estimated variance and the empirical variance, and is used to determined the accuracy of the variance estimator obtained by the multivariate delta method. The accuracy of the interval estimator is determined by calculating the empirical coverage rate (CR) of 95\% confidence interval across 5,000 replications. The simulation results for the point, variance and interval estimates of NIE and MP under the four data types are shown in Table \ref{tab:case1} (Case \#1), Web Table 1 (Case \#2), Web Table 2 (Case \#3) and Web Table 3 (Case \#4), respectively. Detailed findings from the simulation study are reported below.

\subsection{Estimation of NIE}

When the outcome is continuous, the point estimates of NIE generally had minimal bias for sample sizes equal or greater than 500, for all values of TE and MP considered. When the sample was as small as 150, the NIE point estimates had relatively small bias (percent bias within $\pm 10\%$) as long as $\text{MP} \geq 0.2$ and $\text{TE} \geq 0.5$. With binary outcome, however, the point estimates did not show sufficiently small percent bias until the sample size was 1,000, as shown in Web Table 2 (Case \#3) and Web Table 3 (Case \#4). When the outcome was relatively rare and sample size was not large, there was not enough data to accurately estimate the NIE by the product method. In Case \#3 (Web Table 2), we also observed that the percent bias of $\widehat{\text{NIE}}^{(a)}$ was notable  when TE=$\log(2)$ and MP=$0.5$ even when the sample size was 20,000, indicating that with a rare outcome, bias persisted when TE and MP were also large.

The variance ratio of the NIE under all data types and sample sizes were very close to 1, indicating that the variance estimator derived by the multivariate delta method was accurate. Compared to the multivariate delta method, the bootstrap provided more accurate NIE interval estimates 
especially when the sample size is small. When the outcome was continuous, the bootstrap coverage rates were close to or greater than 95\% even when the sample size was 150. In general, both the bootstrap and the multivariate delta method had confidence interval coverage rates very close to the nominal value when the sample size $\geq$ 500, except for the scenario when both TE and MP were large in Case \#3, in which case both methods exhibited lower coverage rates than nominal as sample size increased. This is because the rare outcome assumption may be inadequate and so the resulting bias of  $\widehat{\text{NIE}}^{(a)}$ are more pronounced with a larger sample size. Overall, these findings suggested that the multivariate delta method is a valid approach for estimating the variance and confidence intervals for the NIE when the sample size is at least 500 with a continuous outcome. 

For the cases of binary outcome, we also evaluated the performance of the NIE estimates based on the exact expressions (i.e., $\widehat{\text{NIE}}$). The results are shown in Web Tables 2 and 3 for Cases \#3 and \#4, respectively. As long as the sample sizes are greater than or equal to 1,000, the estimator based on the exact expressions consistently carried small percent bias and nominal coverage rate.

\subsection{Estimation of MP}

In contrast to the NIE estimates, the point estimates of MP often diverged from the true MP values with smaller sample sizes. When the outcome was binary, the percent bias in MP was usually larger than 30\% for sample sizes $\leq$ 1,000, and sometimes even larger than 70\% for sample size $\leq$ 500, as shown in Web Table 2 (Case \#3) and Web Table 3 (Case \#4). The MP point estimates were more stable with a continuous outcome, in which case the bias is negligible when the sample size was at least 500. For the binary outcome scenarios, the MP percent bias was not close to 0 until the sample size was 5,000 or the number of cases $\geq$ 200. For all data types, the percent bias of $\widehat{\text{MP}}$ appears larger when the TE is small. For the same TE, a smaller percent bias occurred when the MP was larger.


The variance estimators of $\widehat{\text{MP}}$ obtained from the multivariate delta method were smaller than their corresponding empirical variances, and this phenomenon became more noticeable with small sample sizes and a small TE. In general, the multivariate delta method provided accurate variance estimators when the sample size was at least 500 for the  continuous outcome scenarios. When the outcome was binary, the multivariate delta method only provided accurate variance estimators when the sample size was at least 5000 and TE $\geq \log(1.5)$.

Similar to the NIE interval estimates, the bootstrap provided more accurate MP interval estimates than the multivariate delta method, especially with smaller sample sizes, because the distribution of $\widehat{\text{MP}}$ with small sample sizes can deviate from normality, which was assumed in the multivariate delta method. For all data types and sample sizes, the bootstrap approach generally had coverage rates higher than 95\% and began to provide close to nominal coverage rates for sample size of 500 for scenarios  with continuous outcome and 5,000 for the scenarios with binary outcome. On the other hand, the multivariate delta method did not provide accurate confidence intervals for smaller sample sizes, and its coverage rates sometimes dropped below 80\% in some settings with a binary outcome. With a continuous outcome, a sample size of 1,000 was needed for the multivariate delta method to provide satisfactory interval estimates, whereas the sample size requirement for the binary outcome scenarios was substantially larger. As shown in Web Table 2 (Case \#3) and Web Table 3 (Case \#4), the multivariate delta method provided satisfactory coverage rates when $\text{TE} \geq \log(1.5)$ and sample size $\geq 5,000$ with a binary outcome. With a the small TE (TE$=\log(1.2)$), a sample size of at least 20,000 will be needed for the multivariate delta method to provide MP interval estimates with nominal coverage. The performance of the delta method was also sensitive to the magnitude of TE. When the TE was small, the MP confidence interval tended to be much wider than it should be when the MP was also small, but tends to be narrower than it should be for larger MP.



\subsection{The Impact of Outcome Prevalence with a Binary Outcome}\label{sec:comp}

We conducted additional simulations to compare the mediation analyses based on the approximate expressions and exact expressions ($\widehat{\text{NIE}}^{(a)}$ v.s. $\widehat{\text{NIE}}$ and $\widehat{\text{MP}}^{(a)}$ v.s. $\widehat{\text{MP}}$) when the baseline outcome prevalence was varied from 1\% to 50\%. We considered TE$\in \{\log(1.2), \log(2)\}$, MP$\in \{0.1,0.5\}$ and a large sample size of 20,000 to alleviate concerns on small-sample biases.

\begin{figure}[htp]
\centering 
\includegraphics[width = 11 cm]{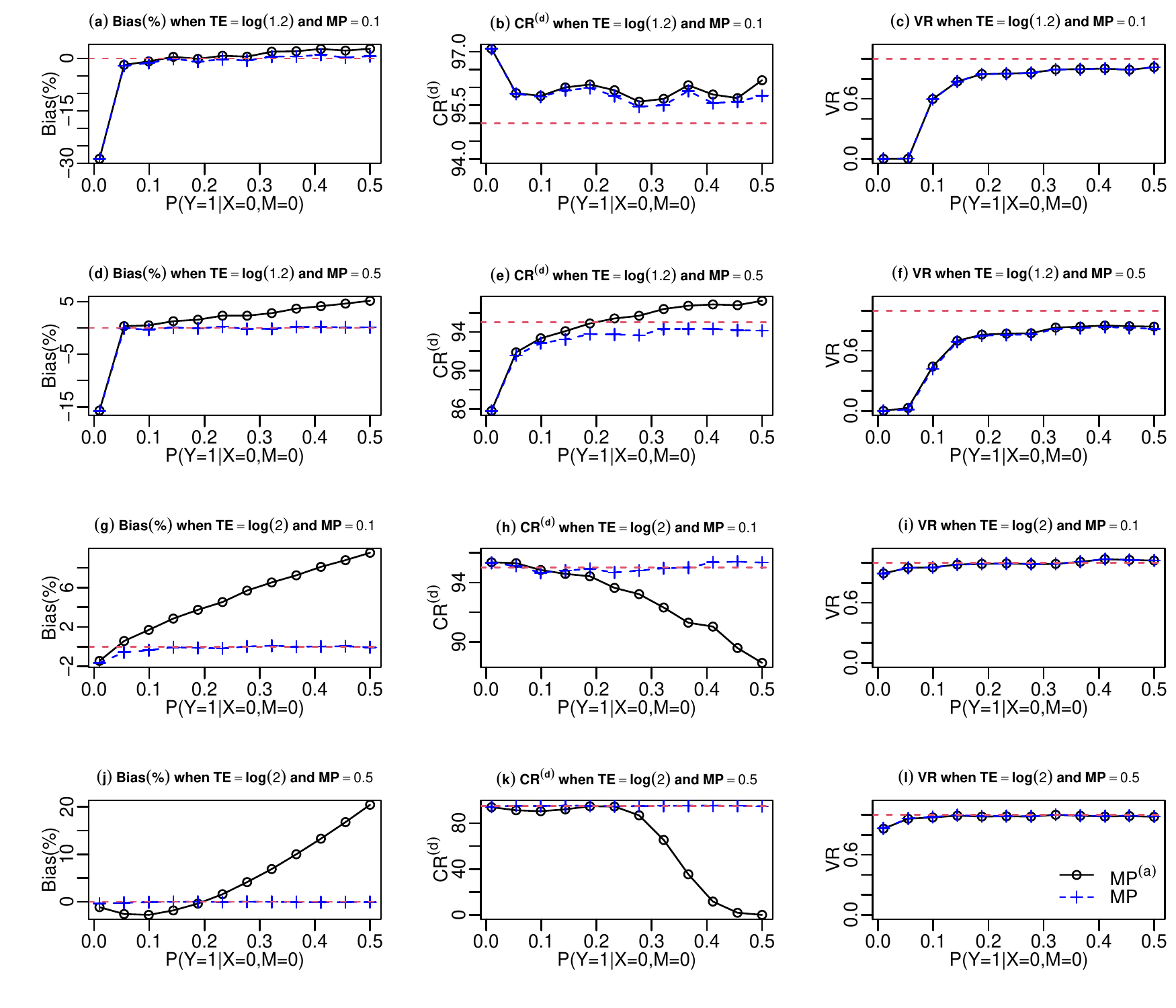}
\caption{Performance of MP$^{(a)}$ estimates (black line) and MP estimates (blue dotted line)  when changing baseline outcome prevalence from 1\% to 50\% in Case \#4, where sample size is 20,000. Bias(\%), CR$^{(d)}$, and VR denote the percent bias, coverage rate by the multivariate delta method, and variance ratio. Upper row: results for TE=log(1.2) and MP=0.1; second row: results for TE=log(1.2) and MP=0.5; third row: results for TE=log(2) and MP=0.1; bottom row: results for TE=log(2) and MP=0.5.}
\label{fig:case3common}
\end{figure}

Figure \ref{fig:case3common} and Web Figure 1 present the results for MP and NIE estimates, respectively, for the binary outcome and binary mediator scenario (i.e., Case \#4).  The estimates based on the exact causal mediation expressions, $\widehat{\text{MP}}$ and $\widehat{\text{NIE}}$, provided accurate point and interval estimates when the outcome prevalence $>1\%$. When $\text{TE}=\log(1.2)$ and the baseline outcome prevalence was 1\%, $\widehat{\text{MP}}$ and $\widehat{\text{NIE}}$  showed significant negative percent bias, as the number of cases (about 200) were quite small. The NIE and MP estimates based on the approximate expressions  did not generally exhibit satisfactory performance when the outcome prevalence was high. Specifically, when TE$> \log(1.2)$, the percent bias of $\widehat{\text{MP}}^{(a)}$ diverged from 0 and the coverage rate of $\widehat{\text{MP}}^{(a)}$ by the delta method substantially decreased from 95\%. Similarly, $\widehat{\text{NIE}}^{(a)}$ was also quite different from its true value when the baseline outcome prevalence $\geq 10\%$ (See Web Figure 1).

With a binary outcome and continuous mediator (Case \#3),  $\widehat{\text{MP}}$ and $\widehat{\text{NIE}}$ provided accurate point and interval estimates among all levels of baseline outcome prevalence, as shown in Web Figures 2 and 3. The estimator $\widehat{\text{MP}}^{(a)}$ also provided very robust point and interval estimates (Web Figure 2) for the common outcome scenarios, where its percent bias was less than $1\%$ among all TEs, MPs and outcome prevalences considered. However, the performance of $\widehat{\text{NIE}}^{(a)}$ was sensitive to the baseline prevalence and the magnitude of MP. For example, when MP=0.5, the percent bias of $\widehat{\text{NIE}}^{(a)}$ increased as baseline outcome prevalence increased, and the confidence interval coverage rates  rapidly declined. When the true MP=0.1, the percent bias of $\widehat{\text{NIE}}^{(a)}$ was negligible over the range of outcome prevalences considered. 

\subsection{Normality Assumption in Case \#3}

In the above simulations, we simulated $M|X$ under a normal distribution, as assumed in deriving the mediation measure expressions in
Case \#3. As a sensitivity analysis, here we evaluate the performance of the product method in Case \#3, when $M$ in fact does not follow the normal distribution. We follow the data generating process for Case \#3, except that we simulated $M=\gamma_0+\gamma_1X+\epsilon$, where $\epsilon$ follows a gamma distribution. Specifically, we let $\epsilon \sim \frac{b-E[b]}{\sqrt{\text{Var}(b)}}$, where $b$ follows a gamma distribution with density $f(b) = \frac{1}{\Gamma(k)\theta^k} b^{k-1} e^{-b/\theta}$ ($b>0$), $k$ and $\theta$ are shape and scale parameters. We subtract $b$ with its expectation $E[b]=k\theta$ and then divide it by its standard deviation $\sqrt{k}\theta$ in order to fix the mean and variance of $\epsilon$ at 0 and 1, matching the first two moments of the standard normal distribution. Here, we chose $k=(2/s)^2$ and $\theta=s/2$ such that the coefficient of skewness of $\epsilon$ was $s$. We choose $s$ to be 1, 1.5 and 2, representing different degrees of skewness, and then still use our exact NIE and MP estimator based on \eqref{eq:NIE3} and \eqref{eq:NDE3}, and the corresponding approximate estimators for analysis. The simulation results are presented in Web Table 4. We also included a scenario that $\epsilon$ follows a standard normal distribution as a benchmark. We observed that both the exact and approximate method are robust with regard to violations of normality assumption, where all the point and variance estimators and confidence interval coverage rates are comparable across the two data generating processes.  This finding suggests that our exact NIE and MP estimators based on \eqref{eq:NIE3} and \eqref{eq:NDE3} are insensitive to moderate skewness of the outcome distribution, even though the derivation assumes normality.





\section{Application to the MaxART study}\label{sec:app}

We performed mediation analysis in the MaxART study \cite{khan2020early}, which is a stepped-wedge cluster randomized trial among HIV-positive participants in Eswatini. The primary objective of the study was to  understand the impact of early access to antiretroviral therapy (EAAA) versus standard of care (SoC). From September 2014 to August 2017, the MaxART Consortium randomly assigned 14 participating clinics in pairs to shift from SoC to EAAA at randomly chosen pre-specified dates. Further details of the design of this MaxART study can be found in \cite{walsh2017impact}. The MaxART study previously found that EAAA improved retention in HIV care \cite{khan2020early}, but the mechanisms underlying the intervention-retention relationship is unknown.  In this illustrative example, we investigated the extent to which the effect of the intervention (SoC v.s. EAAA) on 12-month retention in HIV care was mediated by visit adherence at 6 months. Participants were classified as retained in HIV care for 12 months if, at the end of the 12th month post enrollment, the participant was alive and had not discontinued treatment, where either the last clinic visit was less than 90 days from the end of study or next scheduled visit date was within 30 days from the end of study. In order to obtain 12-month retention, we required (1) the participant's enrollment date to be longer than 12 months from  end of the study and/or (2) if initially receiving SoC treatment, the participant's transition date to EAAA was longer than 12 months from enrollment. Participants who did not meet the above two requirements were excluded. Finally, 1,731 participants were used in our illustrative analysis, with 1,335 individuals retained in care for 12 months and 396 individuals not retained, of whom 1,014 individuals received SoC and 717 individuals received EAAA. Baseline characteristics of participants are given in Web Table 5. For purpose of illustration, we do not addressing clustering of subjects within clinics; we do note, however, minimal degree of clustering has been reported previously in the previous analysis of the MaxART study \cite{khan2020early}. 

The hypothesized mediator considered here was 6-month visit adherence, which measures whether a participant's frequency of clinical visits coincides with the MaxART protocol  over the first 6 months following enrollment. According to the MaxART protocol, participants are expected to have a follow-up visit in every 30 days. It follows that at the end of the 6th month the participants should have completed 6 or more visits. Here, the definition of 6-month visit adherence completion  5 or more clinical visits by the end of the 6th month after enrollment (yes=1, no=0). Based on this definition, 831 participants adhered to the MaxART visit schedule in the first 6 months, whereas 900 participants did not. 

We considered two scenarios for confounding adjustment (i.e., $\bm W$) in the outcome and mediator models. In Scenario I, we only adjusted for the steptime. In Scenario II, we adjusted for all  factors that may have been confounders of the  intervention-retention relationship, visit adherence-retention relationship, or intervention-visit adherence relationship. From clinical knowledge and prior analyses of these data, the comprehensive set of potential confounders included steptime, age at study enrollment ($<20$ yrs, $[20,30)$ yrs, $[30,40)$ yrs, $[40,50)$ yrs, $[50,60)$ yrs, $\geq 60$ yrs), sex, marital status (married, devoiced/widowed, single), education (illiterate/primary, secondary, high school, and tertiary), CD4 counts ($<350$ cells/ul, $[350,500]$ cells/ul, $>500$ cells/ul), WHO stage (I, II, III and IV stages), BMI ($<18.5$, $[18.5,25)$, $[25,30)$, $\geq 30$ $\text{kg/m}^2$), screened for TB symptoms (yes, no), viral load ($<5000$ copies/ml, $[5000,30000]$ copies/ml, $>30000$ copies/ml), treatment support (yes, no), level of clinic (hospital, clinic with maternity ward, clinic without maternity ward), time from HIV tested positive to enrollment ($<1$ yr, 1-3 yrs, $>3$ yrs), and clinic volume (low: $<$ median, high $\geq$ median). For simplicity and consistency with the primary analysis of the MaxART study \cite{khan2020early}, the missing indicator method \cite{groenwold2012missing} was used to account for missing covariates, where missing data was treated as a separate group for each of the confounding variables in the models. 

We first implemented the product method based on the approximate mediation measure expressions assuming a rare outcome. We coded non-retention as 1 and retention as 0. We calculated the NIE, TE and MP comparing EAAA to SoC, conditional on the mode for each model covariate. Although we estimated the mediation measures at the mode of each model covariate, these measures could also be calculated at other values of the covariates as well. Results are given in Table \ref{tab:maxart}. In Scenario I, the steptime-adjusted model, we found that the intervention was protective against 12-month non-retention, with odds ratios of 0.23 and 0.55 for $\widehat{\text{TE}}^{(a)}$ and $\widehat{\text{NIE}}^{(a)}$ respectively. Because the 95\% confidence intervals, either by the delta method or bootstrap, for both parameters excluded the null, we conclude that both effects were significantly different from zero. The steptime-adjusted $\widehat{\text{MP}}^{(a)}$  was 40.8\% (95\% CI by delta method: (0.27, 0.54)), implying that over 40\% of the intervention effect was mediated by 6-month visit adherence. In multivariate-adjusted analyses (Scenario II), stronger NIE and TE effects were obtained, corresponding to odds ratios of 0.08 and 0.38, respectively and the multivariate-adjusted MP estimate was also around 40\%. The bootstrap and multivariate confidence intervals were very close in this analysis, although the width of the bootstrap confidence interval was slightly smaller than that using the delta method variance for NIE and TE, but slightly larger than the delta method for MP.

Because the outcome prevalence in the MaxART study is around 23\%, the above analysis based on the rare outcome approximation may be biased, as suggested by our simulation study. Thus, we repeated the mediation analysis using the exact expressions given in Table \ref{tab:expression}. Generally speaking, the results of the mediation analysis accounting for common outcome prevalence were similar to those obtained using the rare outcome approximation. In both the steptime-adjusted model and the multivariate-adjusted model, the adjusted $\widehat{\text{TE}}$ were slightly weaker than previous results assuming a rare outcome (steptime-adjusted $\widehat{\text{TE}}=0.24$ and multivariate-adjusted $\widehat{\text{TE}}= 0.10$ on the odds ratio scale, compared to 0.23 and 0.08, respectively).  As a result, $\widehat{\text{MP}}$ slightly increased in the adjusted analyses (steptime-adjust ed $\widehat{\text{MP}}=44\%$, multivariate-adjusted $\widehat{\text{MP}}=42\%$), compared to 41\% and 40\%, respectively, with the rare outcome approximation. In summary, for both the results based on and not based on the rare disease assumption, over 40\% of the intervention effect on the 12-month retention in care was mediated by 6-month visit adherence.

\section{Discussion}\label{sec:discussion}

The difference and product methods are two popular approaches for estimating NIE and MP in mediation analysis \cite{vanderweele2015explanation}. While there has been comprehensive empirical evaluations of the difference method \cite{nevo2017estimation}, there were only a few empirical evaluations of the product method as shown in Table \ref{tab:summary}. For this reason, we conducted a comprehensive simulation study to evaluate the performance of $\widehat{\text{NIE}}$ and $\widehat{\text{MP}}$ obtained by the product method under various scenarios likely to be encountered in practice. We also provided the $\widehat{\text{NIE}}$ and $\widehat{\text{MP}}$ estimators without the rare outcome assumption for a binary outcome, and examined extent to which the current approximate mediation analysis was robust to violations of the rare outcome assumption. 

 The estimators investigated in our work has been implemented in an R package \texttt{mediateP} freely available at the Comprehensive R Archive Network (CRAN;  \url{https://cran.r-project.org}); installations and instructions for the R package are given in Appendix A to facilitate their applications. Comparing to the current software for assessing conditional mediation measures (e.g., the SAS and SPSS macros given by \cite{valeri2013mediation} and the SAS macro and R package \texttt{GEEmediate} given by \cite{nevo2017estimation}), the \texttt{mediateP} package gives the exact NIE and MP estimates without rare outcome assumption, when a binary outcome is modeled by a logistic regression. 

We demonstrated that  $\widehat{\text{NIE}}$ had very little bias and the variance estimate for $\widehat{\text{NIE}}$ were quite close to the true values estimated under all scenarios considered from Monte Carlo simulations. In general, the multivariate delta method provided accurate variance estimates and valid interval estimates once the sample size was at least 500, and the bootstrap remained accurate even when the same size was even smaller. We found that larger sample sizes were needed to obtain valid MP point and interval estimates. Specifically, when the outcome was continuous, a sample size of 500 was required for valid point and interval estimates. In the binary outcome scenarios with a rare outcome, a sample size of 5000, and 200 cases or more, were required to obtain satisfactory MP point estimate and bootstrap interval estimates. We observed that the multivariate delta method provided valid MP confidence intervals when sample size $\geq$ 20,000 and number of cases $\geq$ 500 in binary outcome scenario. 

We confirmed that the bootstrap method provided better interval estimates compared the multivariate delta method in smaller sample size scenarios, as may be found in some social science applications. However, the bootstrap method requires substantially more computational time to fit the mediation models   in order to obtain the empirical $\widehat{\text{NIE}}$  or $\widehat{\text{MP}}$  distribution, which may be computationally burdensome in large epidemiological cohort studies. While we recommend bootstrap for the interval estimation with  small sample size, when the sample size is larger, sample size $\geq$ 500 for the continuous outcome scenarios and sample size $\geq$ 20,000 and number of cases $\geq$ 500 for studies of binary outcome, we recommend the multivariate delta method for obtaining valid and  computationally efficient confidence intervals. To facilitate application, our R package \texttt{mediateP} implements both variance estimators. 

In addition, our simulation study also showed that the accuracy of $\widehat{\text{MP}}$ also depends on the effect size of the TE. When the sample size is too small, a smaller TE is associated with a more biased MP point estimates and interval estimates with under-coverage, especially in binary outcome scenarios. In many epidemiological studies, when there is reason to delieve that the NIE or NDE is not close to zero, a relatively smaller sample size may be adequate for obtaining valid point and interval estimates of the MP. For example, when the outcome is binary, we suggest that a sample size $\geq 20,000$ and number of cases $\geq 500$ is needed for the multivariate delta method to obtain satisfactory MP confidence intervals with close to nominal coverage rate. If there is reason to believe  that the TE is not too small (TE$\geq \log(1.5)$), we found that the product method could accurately estimate MP with a sample size of at least 5000 and number of cases at least 150.

In the binary outcome scenarios (Cases \#3 and \#4), expressions of $\text{NIE}^{(a)}$ and $\text{MP}^{(a)}$ defined on a log odds ratio scale have been commonly used in biomedical and epidemiological studies \cite{agerbo2015polygenic,dadvand2014residential,interact2013link}. Those expressions can be extended to include a log link function in the outcome model \eqref{outcome1} and bypass the rare outcome assumption, and the corresponding mediation measure expressions can be defined on a log risk ratio scale. Based on the logistic outcome model, our simulation study suggests that, when the  outcome prevalence was less than 5\%, the rare outcome assumption worked well for $\widehat{\text{NIE}}^{(a)}$ and $\widehat{\text{MP}}^{(a)}$. When the outcome prevalence $\geq$ 5\% and with a binary mediator (Case \#4), the percent bias of $\widehat{\text{NIE}}^{(a)}$ and $\widehat{\text{MP}}^{(a)}$ can be substantial and the exact expressions are recommended. However, with a  binary outcome and continuous mediator (Case \#3), we found that $\widehat{\text{MP}}^{(a)}$ always provides satisfactory point and interval estimates even when the outcome was common.


Our simulation study has several limitations and future work is needed to supplement the conclusions in this manuscript. For simplicity, our simulation study did not include confounders. However, because epidemiological studies need to adjust for all potential  confounding factors to obtain valid results, future research is needed to examine the product method in the presence of multiple confounders. In addition, when the outcome is binary, the mediation measures can be defined on the odds ratio scale \cite{vanderweele2015explanation}. Web Appendix D shows the relationship between mediation measures defined on a log odds ratio scale and odds ratio scale. In summary, we have found that asymptotic inference performs well for the product method in sample sizes typically found in epidemiology and public health settings. In addition, for common binary outcomes, exact expressions are needed to obtain unbiased estimates and strategies for point and variance estimation have been provided here.


\section*{Web Material}
The Web Material, including Web Appendices, Figures, and Tables not shown in the manuscript, is available at \url{https://doi.org/10.6084/m9.figshare.15237054}

\begin{acknowledgement}

This work was funded in part by NIH grant DP1ES025459.

\end{acknowledgement}

\bibliographystyle{abbrvnat}
\bibliography{main}

\section*{Appendix A: Instructions for the \texttt{mediateP} package}


The \texttt{mediateP} package calculates the point and interval estimates for the NIE, TE and MP, based on the product method, as described in this paper. The source files for the \texttt{mediateP} package was provided on CRAN \url{https://cran.r-project.org}. 

First, use the following statements to install the \texttt{mediateP} package
\begin{verbatim}
> install.packages("mediateP")
> library("mediateP") 
\end{verbatim}
The main function of the "mediateP" package is \texttt{mediate()}, which provides the mediation analysis results. It can be called with,
\begin{verbatim}
mediate(data, outcome, mediator, exposure, binary.outcome, 
        binary.mediator, covariate.outcome, covariate.mediator, 
        x0, x1, c.outcome, c.mediator, boot, R)
 \end{verbatim}
The function has 14 arguments. These are
\begin{itemize}
\item[\texttt{data=}] \; (Required) The name of the dataset.

\item[\texttt{outcome=}] \; (Required) Name of the outcome variable, which should be either a continuous or binary datatype.

\item[\texttt{mediator=}] \; (Required) Name of the mediator variable, which should be either a continuous or binary datatype.

\item[\texttt{exposure=}] \; (Required) Name of the exposure variable, which should be either a continuous or binary datatype.

\item[\texttt{binary.outcome=}] \; (Required) If the outcome is binary, set to 1. If the outcome is continuous, set to 0. The default value is 0.

\item[\texttt{binary.mediator=}] \; (Required) If the mediator is binary, set to 1. If the mediator is continuous, set to 0. The default value is 0.

\item[\texttt{covariate.outcome=}] \; A vector of names showing the confounding variables used in the outcome model. The default value is \texttt{NULL}, which represents no confounding variables.  We only accepted continuous and binary confounding variables, if one confounding variable is categorical, please set it to a series of binary variables in advance.

\item[\texttt{covariate.mediator=}] \; A vector of names showing the confounding variables used in the mediator model. The default value is \texttt{NULL}, which represents no confounding variables. We only accepted continuous and binary confounding variables, if one confounding variable is categorical, please set it to a series of binary variables in advance.

\item[\texttt{x0=}] \; (Required) The baseline exposure level (i.e., $x^*$). The default value is 0.

\item[\texttt{x1=}] \; (Required) The new exposure level (i.e., $x$). The default value is 1.

\item[\texttt{c.outcome=}] \; A vector of numbers representing the conditional level of the confounding variables in the outcome model. The default value is a vector of 0.

\item[\texttt{c.mediator=}] \; A vector of numbers representing the conditional level of the confounding variables in the mediator model. The default value is a vector of 0.

\item[\texttt{boot=}] \; If a percentile bootstrap confidence interval needed to be added, set to 1. Otherwise, set to 0. The default value is 0.

\item[\texttt{R=}] \; (Required if \texttt{boot=1}) The number of replications when apply the percentile bootstrap method to calculate the confidence interval. The default value is 2,000.

\end{itemize}

We now illustrate the usage of the \texttt{mediate} function. First, using the following statements to simulate a dataset
\begin{verbatim}
> C1=rnorm(2000)>0 # Confounder 1
> C2=rnorm(2000) # Confounder 2
> X = rnorm(2000) # exposure
> M= as.numeric(runif(2000)< 1/(1+exp(0-0.9*X+0.1*C1))) # mediator
> # outcome
> Y= as.numeric(runif(2000)< 1/(1+exp(-(-2+0.5*X+0.5*M+0.2*C1+0.2*C2))))
> mydata=as.data.frame(cbind(Y,M,X,C1,C2)) # summarize into a dataset
 \end{verbatim}
This dataset, named \texttt{mydata}, includes a continuous exposure (\texttt{X}), a binary mediator (\texttt{M}), a binary outcome (\texttt{Y}), as well as two confounding variables (\texttt{C1} and \texttt{C2}). \texttt{mydata} has 2,000 observations, where the first 6 observations are shown as follows
 \begin{verbatim}
> head(mydata)
  Y M          X C1          C2
1 0 0 -1.1346302  0 -0.88614959
2 0 1  0.7645571  1 -1.92225490
3 0 1  0.5707101  0  1.61970074
4 0 0 -1.3516939  1  0.51926990
5 0 1 -2.0298855  1 -0.05584993
6 0 0  0.5904787  0  0.69641761
 \end{verbatim}
 
 We conducted a mediation analysis using \texttt{mediate()}. In the outcome model, we adjusted for \texttt{C1} and \texttt{C2}. In the mediator model, we only adjusted for \texttt{C1}. We calculated the NIE, TE and MP for exposure in change from 0 to 1, conditional on \texttt{C1=0} and \texttt{C2=1}, as follows
 
\begin{verbatim}
> result=mediate(data=mydata, outcome="Y", mediator="M", exposure="X",
                 binary.outcome=1, binary.mediator=1,
                 covariate.outcome=c("C1","C2"), 
                 covariate.mediator=c("C1"),
                 x0=0, x1=1, c.outcome=c(0,1), c.mediator=c(0),
                 boot=1, R=2000)
 \end{verbatim}
 
Finally, we got the following mediation analysis output
\begin{scriptsize}
 \begin{verbatim}
> print.mediateP(result)
                      Point (S.E.)  95% CI by Delta Approach  95% CI by Bootstrap
NIE: Approximate   0.1069 (0.0224)           (0.0631,0.1508)      (0.0632,0.1510)
NIE:       Exact   0.4568 (0.0645)           (0.3304,0.5832)      (0.3364,0.5840)
TE:  Approximate   0.2341 (0.0588)           (0.1190,0.3493)      (0.1300,0.3710)
TE:        Exact   0.1118 (0.0246)           (0.0635,0.1601)      (0.0647,0.1614)
MP:  Approximate   0.4568 (0.0638)           (0.3318,0.5818)      (0.3402,0.5827)
MP:        Exact   0.2447 (0.0631)           (0.1211,0.3683)      (0.1333,0.3901) \end{verbatim}
 \end{scriptsize}
 
More illustrative examples under other datatypes can be found by using the syntax \texttt{help(mediate)}.

 \newpage

\begin{table}[]
\renewcommand\arraystretch{1.7}
\caption{ A comparison of the current work with several previous literature evaluating the empirical performance of the product method in mediation analysis under four data types: Case \#1, continuous outcome and continuous mediator; Case \#2, continuous outcome and binary mediator; Case \#3, binary outcome and continuous mediator; and Case \#4, binary outcome and binary mediator. }\label{tab:summary}
 \scalebox{0.6}[0.6]{%
\begin{tabular}{@{} p{1in}<{\centering}cccc cccc c cccccc @{}}
\hline
& \multicolumn{7}{c}{Natural Indirect Effect} & \multicolumn{7}{c}{Mediation Proportion} \\
\cmidrule(l{3pt}r{3pt}){2-8} \cmidrule(l{3pt}r{3pt}){9-15} 
Literatures                              & \multirow{2}{*}{Case \#1} & \multirow{2}{*}{Case \#2}  & \multicolumn{3}{c}{Case \#3} & \multicolumn{2}{c}{Case \#4} & \multirow{2}{*}{Case \#1} & \multirow{2}{*}{Case \#2} & \multicolumn{3}{c}{Case \#3} & \multicolumn{2}{c}{Case \#4} \\ 
\cmidrule(l{3pt}r{3pt}){4-6} \cmidrule(l{3pt}r{3pt}){7-8}  \cmidrule(l{3pt}r{3pt}){11-13} \cmidrule(l{3pt}r{3pt}){14-15}
& & & Approx. & Exact & P.A. & Approx. & Exact & & & Approx. & Exact & P.A. & Approx. & Exact \\
\hline
Current work                             & B.V.I & B.V.I & B.V.I      & B.V.I      &       & B.V.I      & B.V.I  & B.V.I  & B.V.I      & B.V.I      & B.V.I  &    & B.V.I & B.V.I      \\
Barfield et al (2017)                     & T &   T  & T      &   &   & T      &       &      &         \\
Biesanz, Falk, and Savalei (2010)        & I.T &     &      &          &          &          &      &            \\
Fritz and MacKinnon (2007)               & T &     &      &          &          &          &      &           \\
Gaynor et al. (2019) & & & &  & B.I & & B.I & & & & & &  \\
MacKinnon, Warsi, and Dwyer (1995)       & B.V &  &       &          &          &    &      & B.V  &       \\
MacKinnon et al. (2002)                  & T &     &       &          &          &          &      &            \\
MacKinnon, Lockwood, and Williams (2004) & I &     &       &          &          &          &      &               \\
Rijnhart et al. (2019)                   &  &  &      B.V    &        &  &  B.V     &     &  &  & B.V  & &  &   B.V          \\ 
Samoilenko, Blais, and Lefebvre (2018) & & & & & & B.V.I & B.V.I\\
\hline
\end{tabular}}
\begin{tablenotes}
     \item[1] Note: The \texttt{B}, \texttt{V}, \texttt{I}, and \texttt{T} denote the bias, variance, confidence interval, and hypothesis testing, respectively. If one of those appears in one cell, it indicates that this operating characteristic has been covered in this literature. In Cases \#3 and \#4, \texttt{Approx.}, \texttt{Exact}, and \texttt{P.A.} denote the approximate expression, exact expression, and the probit approximation expression, respectively (See Table 2 for their specific formulas). 
   \end{tablenotes}
\end{table}

\newcommand{\PXX}{ e^{\beta_0+\beta_1x+\bm \beta_3^Tw}(1+e^{\beta_0+\beta_1x+\beta_2+\bm \beta_3^Tw}) + e^{\gamma_0+\gamma_1x+\bm \gamma_2^Tw}e^{\beta_0+\beta_1x+\beta_2+\bm \beta_3^Tw}(1+e^{\beta_0+\beta_1x+\bm \beta_3^T w})  }
\newcommand{\PSS}{ e^{\beta_0+\beta_1x+\bm \beta_3^Tw}(1+e^{\beta_0+\beta_1x+\beta_2+\bm \beta_3^Tw}) + e^{\gamma_0+\gamma_1x^*+\bm \gamma_2^Tw}e^{\beta_0+\beta_1x+\beta_2+\bm \beta_3^Tw}(1+e^{\beta_0+\beta_1x+\bm \beta_3^T w})  }
\newcommand{\PQQ}{ e^{\beta_0+\beta_1x^*+\bm \beta_3^Tw}(1+e^{\beta_0+\beta_1x^*+\beta_2+\bm \beta_3^Tw}) + e^{\gamma_0+\gamma_1x^*+\bm \gamma_2^Tw}e^{\beta_0+\beta_1x^*+\beta_2+\bm \beta_3^Tw}(1+e^{\beta_0+\beta_1x^*+\bm \beta_3^T w})  }

\newcommand{\MPXX}{ 1+e^{\beta_0+\beta_1x+\beta_2+\bm \beta_3^Tw}+e^{\gamma_0 +\gamma_1x+\bm \gamma_2^Tw}(1 +e^{\beta_0+\beta_1x+\bm \beta_3^Tw})  }
\newcommand{\MPSS}{ 1+e^{\beta_0+\beta_1x+\beta_2+\bm \beta_3^Tw}+e^{\gamma_0 +\gamma_1x^*+\bm \gamma_2^Tw}(1+e^{\beta_0+\beta_1x+\bm \beta_3^Tw})  }
\newcommand{\MPQQ}{ 1+e^{\beta_0+\beta_1x^*+\beta_2+\bm \beta_3^Tw}+e^{\gamma_0 +\gamma_1x^*+\bm \gamma_2^Tw} (1+e^{\beta_0+\beta_1x^*+\bm \beta_3^Tw})  }

\begin{sidewaystable}
\renewcommand\arraystretch{2.9}
\centering
\caption{Expressions of mediation measures under four different datatypes of the outcome and mediator. Case \#1, continuous outcome and continuous mediator; Case \#2, continuous outcome and binary mediator; Case \#3, binary outcome and continuous mediator; and Case \#4, binary outcome and binary mediator.}\label{tab:expression}
 \scalebox{0.67}[0.67]{%
\begin{tabular}{@{} p{0.2in}<{\centering}p{0.4in}<{\centering} | p{3.1in}<{\centering} |  p{4.1in}<{\centering}| p{0.5in}<{\centering} @{}|p{0.9in}<{\centering} @{}}
\hline
  \multicolumn{2}{c|}{Datatypes}       & NIE & NDE & Depend on $\bm W$ & Reference(s)  \\
 \hline
  \multicolumn{2}{c|}{Case \#1}         &  $\beta_2\gamma_1(x-x^*)$   &  $\beta_1(x-x^*)$ & No & \cite{valeri2013mediation}    \\
\hline
  \multicolumn{2}{c|}{Case \#2}                &  $ \beta_2\left\{\frac{\kappa(x,\bm w)-\kappa(x^*,\bm w)}{(1+ \kappa(x,\bm w))(1+ \kappa(x^*,\bm w))}\right\}$   & $\beta_1(x-x^*)$  & Yes &  \cite{barfield2017testing}   \\
\hline
\multicolumn{1}{l|}{}  &  Approx. &  $\beta_2\gamma_1(x-x^*)$  &  $\beta_1(x-x^*)$ & No & \cite{vanderweele2010odds}   \\
\cline{2-6}
\multicolumn{1}{l|}{Case \#3}   &  Exact &   $\log\left\{\frac{ \int_m  \exp(\beta_2 m)\tau(x,x,m,\bm w) \text{d}m  }{  \int_m  \tau(x,x,m,\bm w) \text{d}m  }\right\} - \log\left\{\frac{ \int_m  \exp(\beta_2 m)\tau(x^*,x,m,\bm w)\text{d}m  }{  \int_m \tau(x^*,x,m,\bm w) \text{d}m  }\right\}$  &  $\beta_1(x-x^*)+\log\left\{\frac{ \int_m  \exp(\beta_2 m)\tau(x^*,x,m,\bm w)\text{d}m  }{  \int_m \tau(x^*,x,m,\bm w) \text{d}m  }\right\}-\log\left\{\frac{ \int_m  \exp(\beta_2 m)\tau(x^*,x^*,m,\bm w)\text{d}m  }{  \int_m \tau(x^*,x^*,m,\bm w) \text{d}m  }\right\}$  & Yes & Web Appendix A    \\
\cline{2-6}
\multicolumn{1}{l|}{}  &  Probit Approx.  & $\text{logit}\left(\Phi\left\{\frac{s\beta_0+s\beta_1x+s\beta_3^T \bm w + s\beta_2(\gamma_0+\gamma_1x+\gamma_2^T \bm w)}{\sqrt{1+s^2\beta_2^2\sigma^2}}\right\}\right) - \text{logit}\left(\Phi\left\{\frac{s\beta_0+s\beta_1x+s\beta_3^T \bm w + s\beta_2(\gamma_0+\gamma_1x^*+\gamma_2^T \bm w)}{\sqrt{1+s^2\beta_2^2\sigma^2}}\right\}\right)$  & $\text{logit}\left(\Phi\left\{\frac{s\beta_0+s\beta_1x+s\beta_3^T \bm w + s\beta_2(\gamma_0+\gamma_1x^*+\gamma_2^T \bm w)}{\sqrt{1+s^2\beta_2^2\sigma^2}}\right\}\right) - \text{logit}\left(\Phi\left\{\frac{s\beta_0+s\beta_1x^*+s\beta_3^T \bm w + s\beta_2(\gamma_0+\gamma_1x^*+\gamma_2^T \bm w)}{\sqrt{1+s^2\beta_2^2\sigma^2}}\right\}\right)$ & Yes & \cite{gaynor2019mediation}   \\
\hline
\multicolumn{1}{l|}{Case \#4}   &   Approx.    &     $\beta_2\left\{\frac{\kappa(x,\bm w)-\kappa(x^*,\bm w)}{(1+ \kappa(x,\bm w))(1+ \kappa(x^*,\bm w))}\right\}$   &  $\beta_1(x-x^*)$ & Yes & \cite{vanderweele2015explanation}       \\ 
\cline{2-6}
\multicolumn{1}{l|}{}    &   Exact    &     $\log\left\{\frac{1+e^{\beta_2}\eta(x,\bm w)+\kappa(x^*,\bm w)(1+\eta(x,\bm w))}
{1+e^{\beta_2}\eta(x,\bm w)+\kappa(x,\bm w)(1+\eta(x,\bm w))}\right\} - \log\left\{\frac{1+e^{\beta_2}\eta(x,\bm w)+e^{\beta_2}\kappa(x,\bm w)(1+\eta(x,\bm w))}{1+e^{\beta_2}\eta(x,\bm w)+e^{\beta_2}\kappa(x^*,\bm w)(1+\eta(x,\bm w))}\right\}$   & $\beta_1(x-x^*)+\log\left\{\frac{1+e^{\beta_2}\eta(x^*,\bm w)+\kappa(x^*,\bm w)(1+\eta(x^*,\bm w))}
{1+e^{\beta_2}\eta(x,\bm w)+\kappa(x^*,\bm w)(1+\eta(x,\bm w))}\right\} + \log\left\{\frac{1+e^{\beta_2}\eta(x,\bm w)+e^{\beta_2}\kappa(x^*,\bm w)(1+\eta(x,\bm w))}{1+e^{\beta_2}\eta(x^*,\bm w)+e^{\beta_2}\kappa(x^*,\bm w)(1+\eta(x^*,\bm w))}\right\}$  &  Yes & \cite{gaynor2019mediation} and Web Appendix B   \\ 
\hline
\end{tabular}}
 \begin{tablenotes}
 \footnotesize
 \item[1] Note: NIE and NDE denote the natural indirect effect and natural direct effect, respectively, which are defined for $X$ in change from $x^{*}$ to $x$ conditional on $\bm W=\bm w$, on an identity scale in Cases \#1 and \#2 and a log odds ratio scale in Cases \#3 and \#4.  Given NIE and NDE, the mediation proportion (MP) can be obtained by $\frac{\text{NIE}}{\text{NIE}+\text{NDE}}$. In the probit approximation method, $s=1/1.6$, $\text{logit}(x)=\log(\frac{x}{1-x})$, and $\Phi(.)$ is the cumulative density function for the standard normal distribution. 
   \end{tablenotes}
\end{sidewaystable}

\begin{table}[htbp]
\renewcommand\arraystretch{0.7}
  \centering
  \caption{Simulation results for Case \#1: continuous outcome and continuous mediator.}\label{tab:case1}
 \scalebox{0.75}[0.75]{%
\begin{tabular}{ccc rrrr rrrr}
\toprule
\multicolumn{3}{c}{ } & \multicolumn{4}{c}{$\widehat{\text{NIE}}$} & \multicolumn{4}{c}{$\widehat{\text{MP}}$}  \\
\cmidrule(l{3pt}r{3pt}){4-7} \cmidrule(l{3pt}r{3pt}){8-11} 
N &  MP & TE & Bias(\%) & CR$^{(d)}$ & CR$^{(b)}$ & VR &  Bias(\%) & CR$^{(d)}$ & CR$^{(b)}$ & VR \\
\midrule
& 0.05 & 0.25 & -16.4 & \textbf{99.2} & \textbf{96.4} & 0.998 & -27.5 & \textbf{99.2} & \textbf{98.8} & 0.000\\

 &  & 0.5 & -15.1 & \textbf{96.5} & \textbf{95.7} & 0.992 & -14.1 & \textbf{96.9} & \textbf{96.5} & 0.058\\

 &  & 1 & -10.5 & \textbf{92.2} & 94.7 & 0.993 & -9.8 & \textbf{92.6} & 95.1 & 0.949\\

 & 0.2 & 0.25 & -10.5 & \textbf{92.2} & 94.7 & 0.993 & -19.7 & \textbf{90.6} & \textbf{97.3} & 0.000\\

 &  & 0.5 & -6.2 & \textbf{92.5} & 94.8 & 1.001 & -2.8 & \textbf{92.9} & \textbf{96.4} & 0.092\\

 &  & 1 & -2.1 & 94.8 & 94.9 & 1.012 & -1.7 & \textbf{95.7} & 95.4 & 0.944\\

 & 0.5 & 0.25 & -4.7 & \textbf{93.2} & 94.8 & 1.007 & -14.1 & \textbf{89.1} & \textbf{97.0} & 0.000\\

 &  & 0.5 & -1.6 & 95.1 & 95.0 & 1.012 & -0.4 & \textbf{96.0} & \textbf{97.0} & 0.026\\

\multirow{-9}{*}{150} &  & 1 & -0.1 & 95.5 & 94.9 & 1.013 & 0.2 & \textbf{97.5} & 95.3 & 0.710\\
\cmidrule{1-11}
 & 0.05 & 0.25 & -4.4 & \textbf{96.0} & 94.4 & 0.997 & -3.5 & \textbf{97.8} & \textbf{96.6} & 0.002\\

 &  & 0.5 & -3.5 & 94.7 & \textbf{94.2} & 0.988 & -3.4 & 95.3 & 94.5 & 0.855\\

 &  & 1 & -2.6 & \textbf{93.1} & \textbf{94.2} & 0.983 & -2.8 & \textbf{93.4} & \textbf{94.4} & 0.948\\

 & 0.2 & 0.25 & -2.6 & \textbf{93.1} & \textbf{94.2} & 0.983 & -1.5 & \textbf{92.1} & \textbf{97.0} & 0.001\\

 &  & 0.5 & -1.5 & \textbf{93.6} & 94.5 & 0.982 & -0.8 & \textbf{94.4} & 94.8 & 0.801\\

 &  & 1 & -0.2 & 94.5 & 94.5 & 0.981 & -0.1 & 94.7 & 94.6 & 0.945\\

 & 0.5 & 0.25 & -0.9 & \textbf{94.0} & 94.5 & 0.980 & -0.4 & \textbf{92.3} & \textbf{97.5} & 0.000\\

 &  & 0.5 & -0.1 & 94.7 & \textbf{94.3} & 0.982 & 0.1 & 95.5 & 94.9 & 0.770\\

\multirow{-9}{*}{ 500} &  & 1 & 0.6 & 94.5 & 94.5 & 0.983 & 0.0 & 95.6 & 94.7 & 0.915\\
\cmidrule{1-11}
 & 0.05 & 0.25 & -4.0 & 95.4 & 94.7 & 0.969 & -3.2 & \textbf{97.0} & 95.4 & 0.724\\

 &  & 0.5 & -2.2 & 94.8 & 94.7 & 0.969 & -2.8 & 95.2 & 95.1 & 0.924\\

 &  & 1 & -2.1 & \textbf{94.1} & 94.8 & 0.971 & -1.9 & \textbf{94.0} & 94.7 & 0.967\\

 & 0.2 & 0.25 & -2.1 & \textbf{94.1} & 94.8 & 0.971 & -1.1 & \textbf{93.5} & \textbf{96.3} & 0.580\\

 &  & 0.5 & -1.1 & \textbf{94.2} & 94.6 & 0.981 & -1.2 & \textbf{94.2} & 94.8 & 0.923\\

 &  & 1 & -0.9 & 94.9 & 94.8 & 0.986 & -0.8 & 94.6 & 94.4 & 0.981\\

 & 0.5 & 0.25 & -1.1 & 94.4 & 94.6 & 0.983 & -0.5 & \textbf{92.8} & \textbf{96.6} & 0.441\\

 & & 0.5 & -0.5 & 95.0 & 94.9 & 0.987 & -0.5 & 94.9 & 94.6 & 0.916\\

\multirow{-9}{*}{ 1000} &  & 1 & -0.3 & 94.8 & 94.7 & 0.988 & -0.4 & 95.0 & 94.5 & 0.966\\
\cmidrule{1-11}
 & 0.05 & 0.25 & -0.5 & 94.8 & 94.8 & 0.990 & 0.3 & 95.5 & 94.8 & 0.961\\

 &  & 0.5 & -0.2 & 94.9 & 95.0 & 0.995 & -0.3 & 95.4 & 95.0 & 0.990\\

 &  & 1 & -0.2 & 95.0 & 95.1 & 1.007 & -0.3 & 95.0 & 95.1 & 1.006\\

 & 0.2 & 0.25 & -0.2 & 95.0 & 95.1 & 1.007 & 0.1 & 95.6 & 95.1 & 0.936\\

 &  & 0.5 & -0.1 & 95.0 & 95.2 & 1.027 & -0.1 & 95.1 & 95.3 & 1.004\\

 &  & 1 & 0.0 & 95.3 & 95.2 & 1.038 & 0.0 & 95.1 & 95.1 & 1.031\\

 & 0.5 & 0.25 & 0.0 & 95.3 & 95.5 & 1.032 & 0.1 & 95.0 & 94.7 & 0.920\\

 &  & 0.5 & -0.1 & 95.3 & 95.2 & 1.040 & 0.0 & 95.2 & 95.0 & 0.995\\

\multirow{-9}{*}{ 5000} &  & 1 & 0.0 & 95.4 & 95.2 & 1.041 & 0.0 & 95.3 & 95.4 & 1.022\\
\bottomrule
\end{tabular}}
  \begin{tablenotes}
     \item[1] \small{Note: Bias(\%), CR$^{(d)}$, CR$^{(b)}$, and VR denote the median percent bias, 95\% confidence interval coverage rate of multivariate delta method, 95\% confidence interval coverage rate of percentile bootstrap method, and mediation variance ratio, respectively. The coverage rates outside the 95\% confidence boundary, i.e., $q \pm 1.96\times \sqrt{\frac{q(1-q)}{B}}$, were highlighted in bold, where $q$ denotes the nominal confidence interval threshold (95\%) and $B$ denotes the number of replication (5,000).  The median percent bias was calculated as the median of the ratio of bias to the true value over 5,000 replications, i.e., $\text{Bias(\%)}=median(\frac{\hat{p}-p}{p}) \times 100\%$, where $p$ denotes the true value of the causal mediation measure, and $\hat{p}$ is the point estimate of the simulated causal mediation measure. The median variance ratio is defined by the ratio of median delta-method variance estimators across 5,000 replications to the empirical variance of causal mediation measure estimates from the 5,000 replications.}
   \end{tablenotes}
\end{table}%

\begin{table}[htbp]
\centering
\caption{Mediation analysis of MaxART \cite{khan2020early}. (n=1731)}\label{tab:maxart}
 \scalebox{0.75}[0.75]{%
\begin{tabular}{ccccccc}
 \toprule
Expression & Scenario & Parameter & Point & S.E. & Delta 95\% CI & Bootstrap 95\% CI \\ 
  \hline
 \multirow{6}{*}{Steptime adjusted} & &  NIE$^{(a)}$ & -0.601 & 0.091 & (-0.779,-0.424) & (-0.782,-0.445) \\ 
& Approximate  & TE$^{(a)}$ & -1.472 & 0.226 & (-1.915,-1.030) & (-1.973,-1.031) \\ 
&  & MP$^{(a)}$ & 0.408 & 0.069 & (0.273,0.544) & (0.305,0.559) \\ 

   \cmidrule{2-7}
& & NIE & -0.630 & 0.093 & (-0.813,-0.448) & (-0.816,-0.474) \\ 
& Exact  &TE & -1.444 & 0.213 & (-1.862,-1.027) & (-1.922,-1.023) \\ 
&  & MP & 0.437 & 0.073 & (0.293,0.580) & (0.327,0.597) \\ 
      \cmidrule{1-7}

\multirow{6}{*}{Multivariate adjusted}   & & NIE$^{(a)}$ & -0.972 & 0.121 & (-1.208,-0.735) & (-1.287,-0.775) \\ 
& Approximate & TE$^{(a)}$ & -2.520 & 0.287 & (-3.082,-1.958) & (-3.282,-1.975) \\ 
 & & MP$^{(a)}$ & 0.386 & 0.050 & (0.288,0.483) & (0.292,0.494) \\  

 \cmidrule{2-7}
&  & NIE & -0.970 & 0.120 & (-1.205,-0.735) & (-1.282,-0.772) \\ 
& Exact & TE & -2.316 & 0.267 & (-2.841,-1.792) & (-3.033,-1.819) \\ 
&  & MP & 0.419 & 0.054 & (0.313,0.525) & (0.322,0.539) \\ 
  \bottomrule
\end{tabular}}
\begin{tablenotes}
     \item[1] Note: All the mediation measures, including NIE, TE, and MP, are defined on a log odds ratio scale for the intervention in change from SoC to EAAA, conditional on the most frequent level of the confounding variables. S.E. denotes the standard error of the point estimates, which is calculated by the multivariate delta method. We implemented the delta method and  bootstrap method with 1,000 replications to calculate the 95\% confidence interval (95\% CI) of each mediation measure.
   \end{tablenotes}
\end{table}

\end{document}